\documentclass[a4paper,11pt]{article}
\usepackage{pos}
\usepackage{lineno}
\newcommand{\pp}{\ensuremath{\mathrm {p\kern-0.05em p}}}
\newcommand{\PbPb}{\ensuremath{\mbox{Pb--Pb}}}
\newcommand{\GeVc}{\ensuremath{\mathrm{GeV}\kern-0.05em/\kern-0.02em c}}

\newcommand{\pT}{\ensuremath{p_{\mathrm{T}}}}
\newcommand{\pTsub}{\ensuremath{p_{\mathrm{T,subleading}}}}
\newcommand{\pTlead}{\ensuremath{p_{\mathrm{T,leading}}}}
\newcommand{\kT}{\ensuremath{k_{\mathrm{T}}}}

\newcommand{\pTjet}{\ensuremath{p_{\mathrm{T,\;jet}}}}

\newcommand{\tg}{\ensuremath{\theta_{\mathrm{g}}}}

\newcommand{\rg}{\ensuremath{R_{\mathrm{g}}}}
\newcommand{\zg}{\ensuremath{z_{\mathrm{g}}}}

\newcommand{\RAA}{\ensuremath{R_{\mathrm{AA}}}}

\title{Overview of the latest jet physics results from ALICE}
\ShortTitle{Overview of the latest jet physics results from ALICE}

\author[a,b]{James Mulligan for the ALICE Collaboration}
\emailAdd{james.mulligan@berkeley.edu}
\affiliation[a]{Nuclear Science Division, Lawrence Berkeley National Laboratory, Berkeley, California 94720, USA}
\affiliation[b]{Physics Department, University of California, Berkeley, CA 94720, USA}

\abstract{
We overview recent jet measurements in \pp{} and \PbPb{} collisions with the ALICE detector.
ALICE reconstructs jets at midrapidity using the anti-\kT{} algorithm, including both
charged particle jets from the ALICE tracking detectors, and
full jets from the combination of the ALICE electromagnetic calorimeter with the tracking system.
We focus on inclusive and semi-inclusive jet measurements as well as jet substructure measurements, including a variety of groomed and ungroomed observables.
In \pp{} collisions, these results test pQCD techniques at $\pTjet<140\;\GeVc$ in a low pileup
environment, and can constrain models of non-perturbative effects. 
In \PbPb{} collisions, these measurements test models of jet quenching
in the quark-gluon plasma, and can be used to constrain the properties of high temperature QCD matter. 
}

\FullConference{%
  40th International Conference on High Energy physics - ICHEP2020\\
  July 28 - August 6, 2020\\
  Prague, Czech Republic (virtual meeting)
}

\begin{document}
\maketitle

\section{Introduction}

Jet measurements can be used to study fundamental aspects of QCD
in both \pp{} and \PbPb{} collisions.
In \pp{} collisions, jet measurements test perturbative calculations, 
which is important for our first-principles understanding of QCD, and
inform which observables are under sufficient theoretical control to 
serve as suitable baselines for heavy-ion measurements. 
Jet measurements in \pp{} collisions also constrain nonperturbative effects, such as hadronization. 
In \PbPb{} collisions, jets serve as probes of the quark-gluon plasma (QGP). 
Comparing jet measurements in \PbPb{} collisions to those in \pp{}
collisions allows us to investigate the modifications of jets due to their interaction
with the QGP. 
By comparing models to data, we can constrain medium bulk properties such as
transport coefficients, and potentially
elucidate the nature of the degrees of freedom of the QGP.

In what follows, we highlight a selection of recent results from the ALICE experiment,
with an emphasis on inclusive, semi-inclusive, and jet substructure measurements. 
We reconstruct jets at midrapidity using the anti-\kT{} 
algorithm 
with resolution parameters ranging from $R=0.1$ to $R=0.6$.
These include both ``charged jets''
clustered only from charged particles,
as well as ``full jets'' clustered with both charged and neutral particles.
All presented results are corrected for detector effects (in both \pp{} 
and \PbPb{} collisions) and background fluctuations (in \PbPb{} collisions).

\section{Jet measurements in proton--proton collisions}

\subsection{Inclusive cross-sections}

Inclusive jet measurements at low-\pT{} as a function of $R$ provide tests of the perturbative
and nonperturbative (NP) contributions to the inclusive jet cross-section 
\cite{PhysRevC.101.034911, ppCMS13TeV}.
First-principles calculations of the perturbative cross-section were recently computed at NLO
with resummation of large logarithms \cite{Dasgupta2016, SiJF, JointResummation2018}, and to NNLO 
at fixed order \cite{NNLO, Czakon2019}.
The significance of various terms of the perturbative expansion, as well as the significance of the hadronization and underlying event (UE) effects, are important questions 
for our fundamental understanding of QCD.

\begin{figure}[!ht]
\centering{}
\includegraphics[scale=0.43]{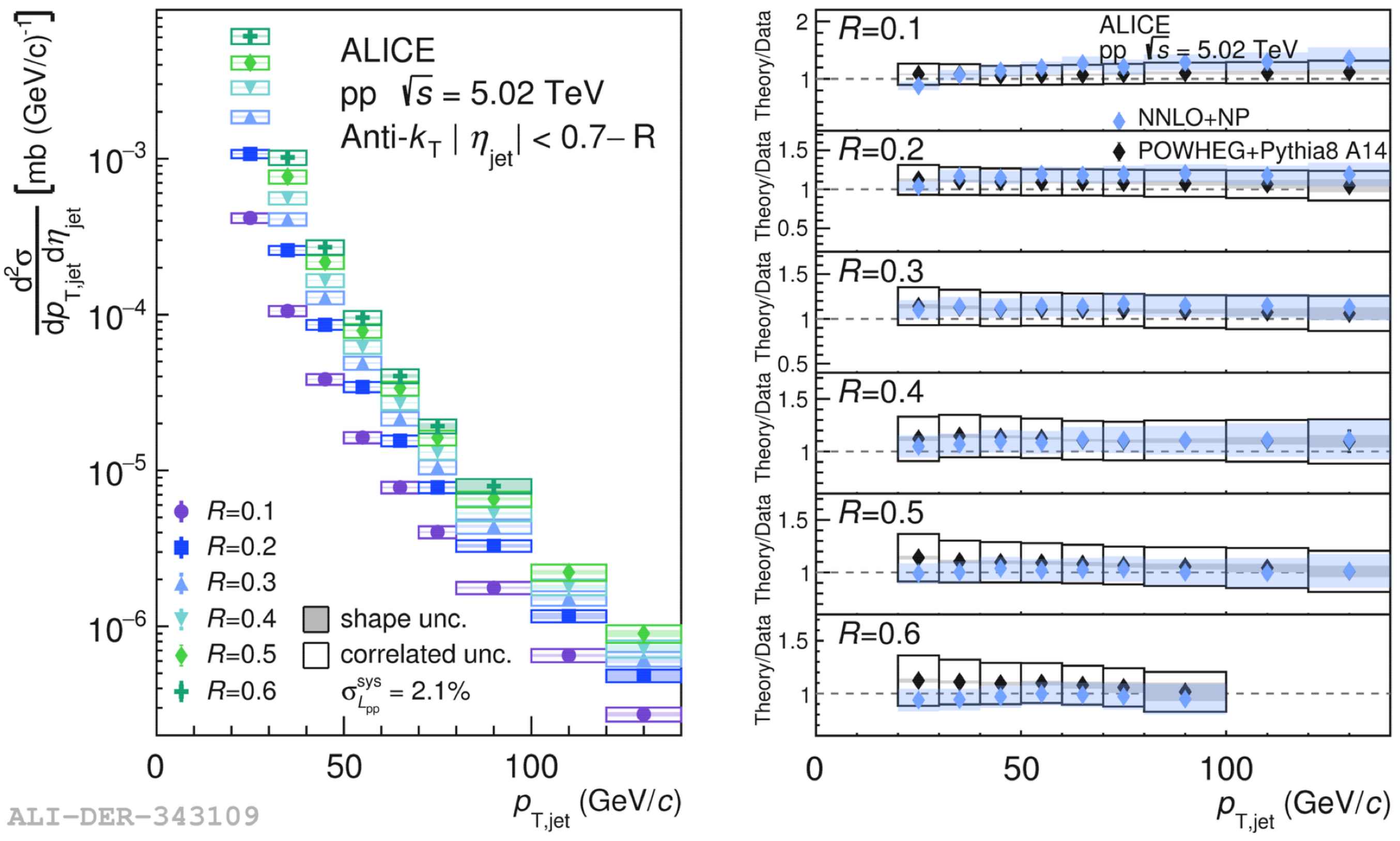}
\caption{Measurements of inclusive jet cross-sections in \pp{} collisions for $R=0.1-0.6$ (left) \cite{PhysRevC.101.034911} and 
comparison to NNLO calculations \cite{NNLO} (with NP corrections) 
and POWHEG+PYTHIA8 (right) \cite{powheg1,pythia}.}
\label{fig:inclusive}
\end{figure}

ALICE recently reported inclusive jet cross-sections for jet resolution parameters $R=0.1-0.6$ over the range $20<\pTjet<140$ \GeVc, shown in Fig. \ref{fig:inclusive} (left) \cite{PhysRevC.101.034911}. 
By covering a large range of $R$ down to low \pT{},
these measurements span a range of perturbative regimes including small-$R$ resummation,
and a wide range of NP effects (from hadronization-dominated at small $R$ to UE-dominated at large $R$),
which can be used to further constrain NP effects in \pp{} collisions.
Figure \ref{fig:inclusive} (right) shows a comparison of these measurements to 
NNLO calculations \cite{NNLO} as well as POWHEG+PYTHIA8 \cite{powheg1,pythia}. 
These predictions are consistent with the data for all $R$ and \pTjet{}, 
and along with other comparisons \cite{PhysRevC.101.034911} demonstrate the importance of NNLO effects and NLL resummations. 
We additionally reported jet cross-section ratios of different $R$,
which allow one to elucidate higher-precision effects of the $R$-dependence of the inclusive jet cross-section.

\subsection{Dynamical grooming}

\begin{figure}[!ht]
\centering{}
\includegraphics[scale=0.3]{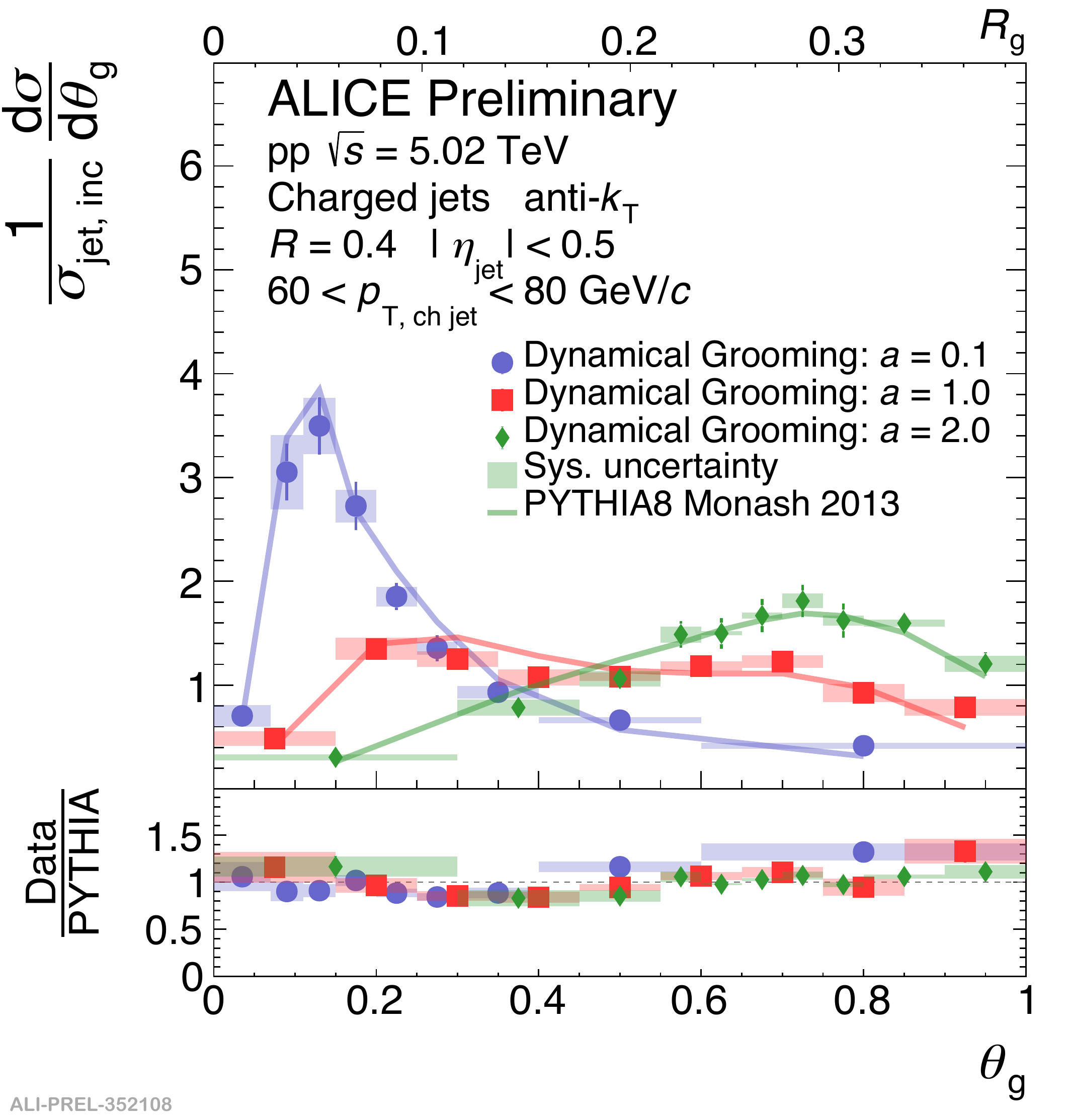}
\includegraphics[scale=0.305]{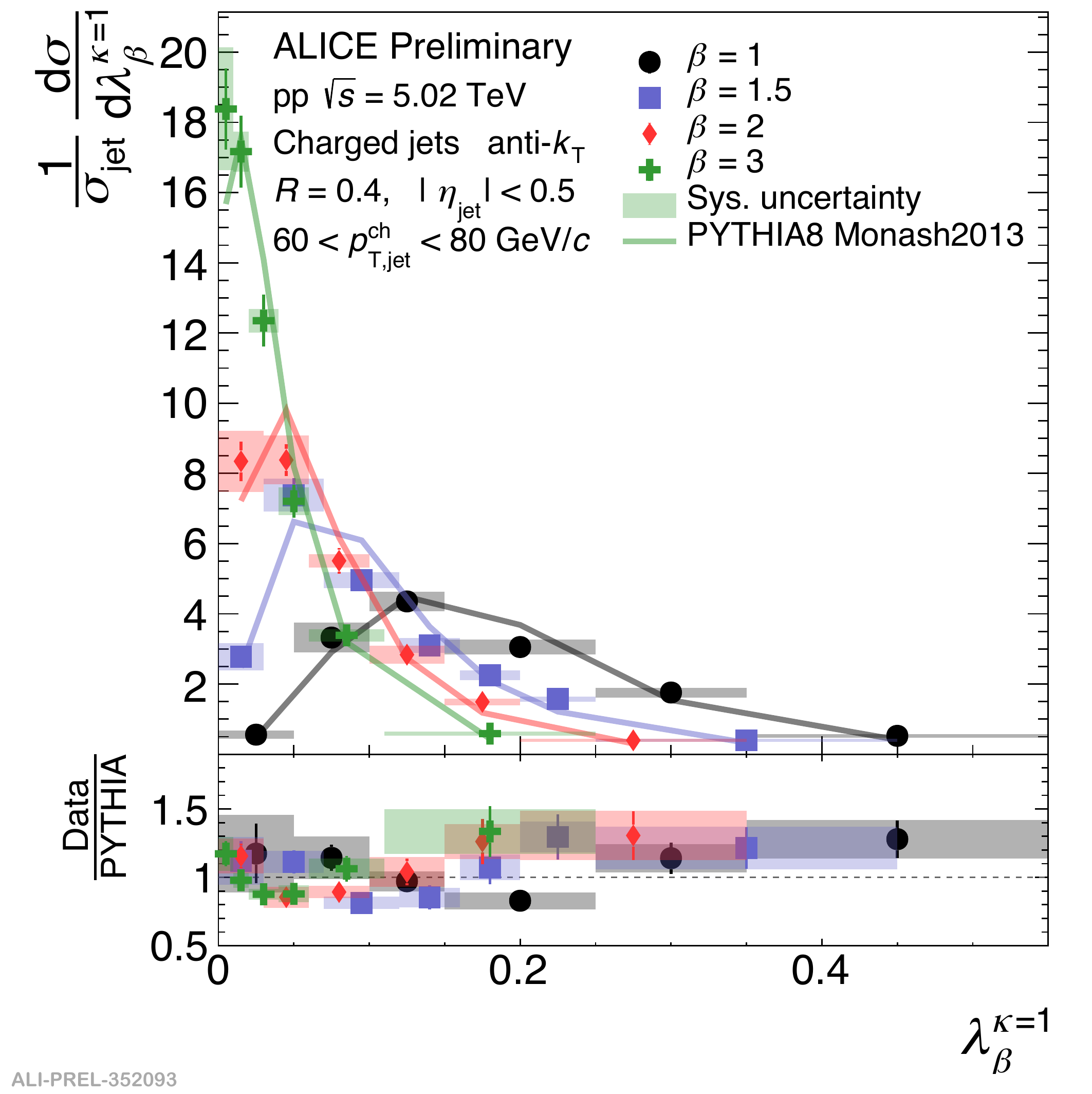}
\caption{Measurements of \tg{} (left) in \pp{} collisions with
Dynamical Grooming \cite{PhysRevD.101.034004} for three values of the
grooming parameter $a$.
Measurements of jet angularities (right) $\lambda_\beta^{\kappa=1}$ in \pp{} collisions with
$R=0.4$ for four values of the
continuous parameter $\beta$. 
Both results are compared to PYTHIA Monash 2013 \cite{pythia}.}
\label{fig:dg-ang}
\end{figure}

Jet grooming techniques are used
reduce non-perturbative effects by selectively removing soft 
large-angle radiation, which allows for well-controlled comparisons of measurements 
to pQCD calculations
\cite{Larkoski_2020, Larkoski:2014wba, PhysRevD.101.034004, Kang:2019prh, PhysRevD.101.052007, PhysRevD.98.092014, STAR2020}.
The Dynamical grooming algorithm \cite{PhysRevD.101.034004,mehtartani2020tagging} identifies a single ``splitting'' by 
re-clustering the constituents of a jet with the Cambridge-Aachen algorithm \cite{Dokshitzer_1997}, 
and traversing the primary Lund plane \cite{Dreyer_2018} to identify the splitting that maximizes:
$
z_i (1-z_i)p_{\rm{T},i} \left( \frac{\Delta R_i}{R} \right)^a,
$
where $z_i$ is the longitudinal momentum fraction of the $i^{\rm{th}}$ splitting,
$\Delta R_i$ is the rapidity-azimuth ($y,\varphi$) separation of the daughters,
and $a$ is a continuous free parameter. 
Since the grooming condition defines a maximum rather than an explicit cut,
every jet returns a tagged splitting.
We focus on the two kinematic observables that characterize the splitting: 
the groomed jet radius, $\tg \equiv \rg/R \equiv \sqrt{\Delta y ^2 + \Delta \varphi ^2}/R$,
and the groomed momentum fraction, $\zg \equiv \pTsub / (\pTlead + \pTsub)$.
Figure \ref{fig:dg-ang} (left) shows the \tg{} distributions in \pp{} collisions
for several values of the grooming parameter $a$. 
For small $a$, the grooming condition favors splittings with symmetric
longitudinal momentum, reflected in the distributions skewing towards
small \tg. As $a$ increases, the grooming condition favors 
splittings with large angular separation, reflected in the distributions 
skewing towards large-\tg.
The results are compared to PYTHIA \cite{pythia}, 
which describes the data well.

\subsection{Ungroomed jet angularities}

The class of infrared and collinear safe jet angularities \cite{Larkoski_2014} is defined
as 
\begin{equation} \label{ang_eqn}
\lambda_\beta^{\kappa} =\sum\limits_{i \in \text{jet}}
\bigg( \frac{p_{\text{T},i}}{p_{\text{T, jet}}} \bigg)^{\kappa}
\bigg( \frac{\Delta R_i}{R} \bigg)^\beta 
\end{equation}
for $\kappa=1$ and $\beta>0$. 
Jet angularities provide a 
flexible way to study QCD in both \pp{} 
and \PbPb{} \cite{Aad_2012, PhysRevD.98.092014, ang2018} collisions due to the ability
to systematically vary the observable definition in a way that is theoretically calculable.
Additionally, jet angularities give sensitivity to the predicted scaling of 
non-perturbative shape functions \cite{Kang_2018, KANG201941}.
Figure \ref{fig:dg-ang} (right) shows the $\lambda_\beta^{\kappa=1}$ distributions in \pp{} collisions
(right) for several values of $\beta$. 
As $\beta$ increases, the distributions skew towards small $\lambda_\beta^{\kappa=1}$, 
since $\Delta R_i/R$ is smaller than unity. 
The distributions become broader for smaller $R$ (not shown here), 
as expected due to the collinear nature of jet fragmentation.
The results are compared to PYTHIA \cite{pythia}, 
which describes the data reasonably well.

\section{Jet measurements in heavy-ion collisions}

\subsection{Inclusive jet \RAA}

\begin{figure}[!ht]
\centering{}
\includegraphics[scale=0.3]{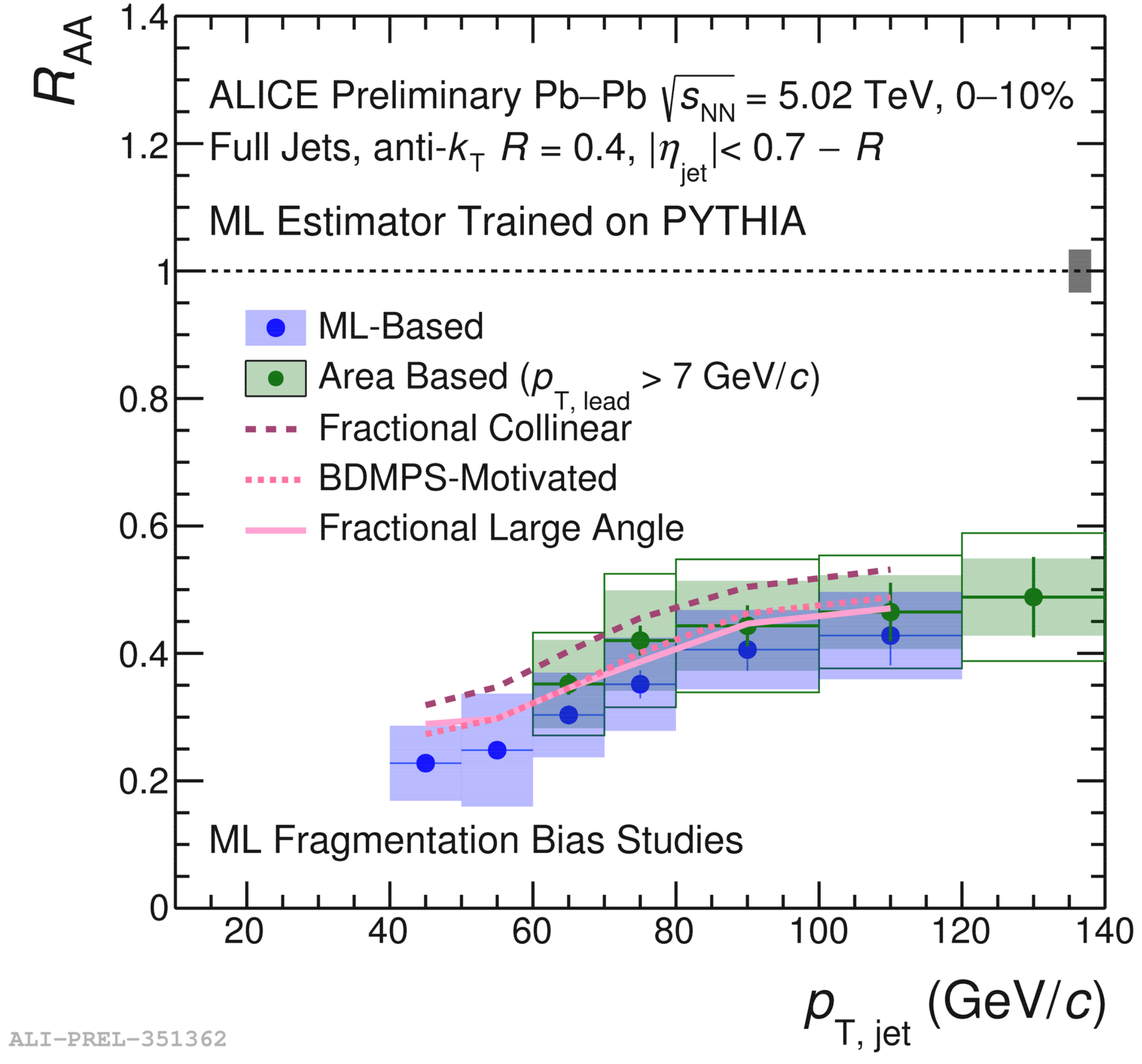}
\qquad
\includegraphics[scale=0.26]{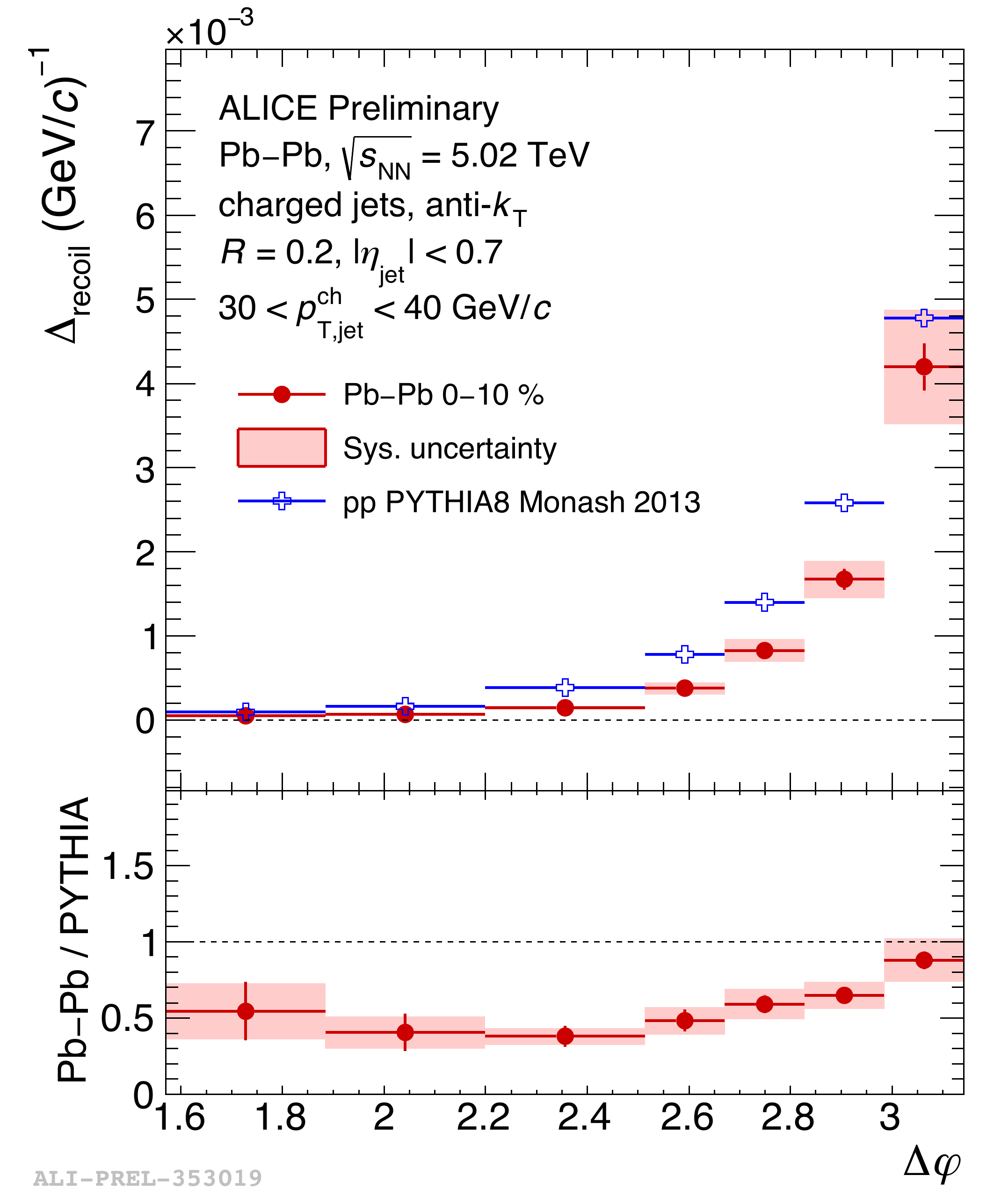}
\caption{Measurements of jet \RAA{} (left) using a PYTHIA-based machine learning method for 
background subtraction \cite{bossi2020inclusive}, and per-trigger semi-inclusive
jet yields (right) as a function of the azimuthal angle difference
$\Delta \varphi$ between trigger hadrons and associated recoiling jets \cite{norman2020jet}. 
}
\label{fig:inclusiveAA}
\end{figure}

The jet \pT{} spectrum in heavy-ion collisions is suppressed relative to that in (appropriately scaled) \pp{} collisions, 
indicating that jets transfer energy to the hot QCD medium (see e.g. \cite{PhysRevC.101.034911}).
Extending measurements to low \pT{} and large $R$ is of particular interest, 
in order to constrain competing effects between
the recovery of out-of-cone radiation and changes in the jet population (see e.g. \cite{Qiu:2019sfj}).
To do this, ALICE has started exploring
the use of machine learning (ML) to estimate the background-subtracted jet \pT{}
on a jet-by-jet basis,
reducing UE fluctuations at the expense of introducing model bias. 
Figure \ref{fig:inclusiveAA} (left) shows an example in which
the ML method extends the inclusive jet measurement
to lower \pT{} compared to traditional techniques \cite{bossi2020inclusive}. 
The ML method is trained using PYTHIA, however, and is based 
on subtracting more \pT{} from jets with a higher fraction of soft particles 
(typically arising from the 
UE) relative to jets with a higher fraction of hard particles (typically arising from jet fragmentation).
This method therefore assumes that jet modification in the QGP (which is known to induce
soft and wide-angle energy flow) does not significantly alter the ML performance,
otherwise the method will lead to uncontrolled errors in the \pT{} reconstruction.
Several simple models of modified fragmentation are shown in Fig. \ref{fig:inclusiveAA} (left),
and show significant model-dependence. These effects remain under investigation.

\subsection{Semi-inclusive hadron-jet correlations}

Semi-inclusive hadron-jet correlations are well-suited to 
statistical background subtraction procedures in heavy-ion collisions, which
allows jet measurements to low \pT{} and large $R$ \cite{hjetPbPb, hjetAuAu}.
By measuring the azimuthal angular separation between  trigger hadrons and associated 
recoiling jets, one can test for large-angle jet deflection in the QGP \cite{D'Eramo2019}
as well as transverse broadening (e.g. \cite{Gyulassy_2019}).
Figure \ref{fig:inclusiveAA} (right) shows the distribution of per-trigger semi-inclusive
yields as a function of the azimuthal angle difference
$\Delta \varphi$ between trigger hadrons and associated recoiling jets \cite{norman2020jet}.
This is the first hadron-jet $\Delta \varphi$ distribution that is fully corrected for
detector and background effects. The measurement shows an overall suppression of 
yields in \PbPb{} collisions relative to PYTHIA, which is typical of medium-induced energy loss.
The measurement also shows a relative narrowing of the $\Delta \varphi$ distribution towards
$\Delta \varphi = \pi$. The origin of this apparent narrowing is unknown (see for example 
\cite{zakharov2020radiative}), and demands further theoretical and experimental study.

\subsection{Soft Drop grooming}

\begin{figure}[!ht]
\centering{}
\includegraphics[scale=0.27]{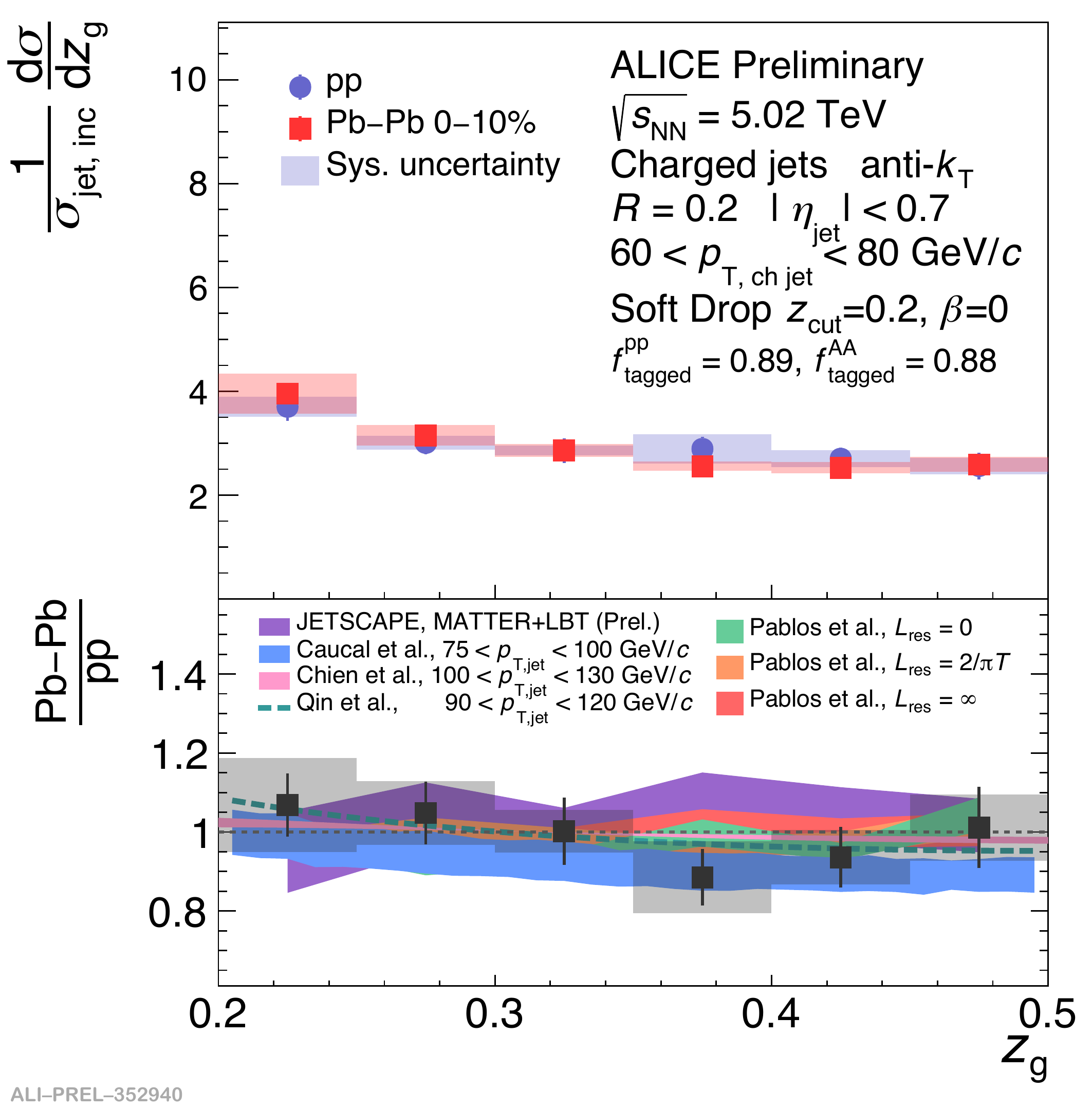}
\includegraphics[scale=0.27]{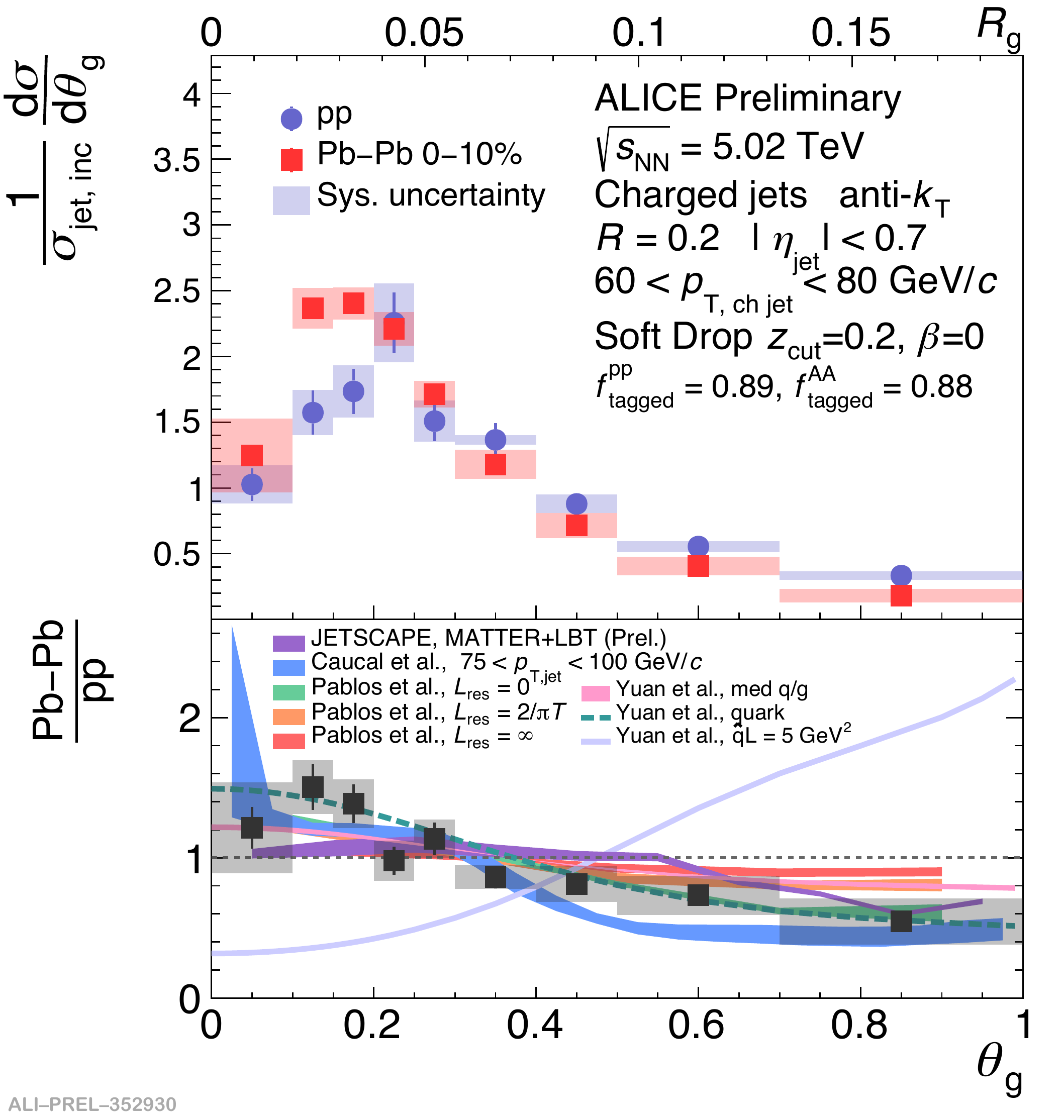}
\caption{Measurements of \zg{} (left) and \tg{} (right) in 0--10\% central \PbPb{} collisions compared to \pp{} collisions for $R=0.2$, along with comparison to several theoretical models \cite{ALICE-PUBLIC-2020-006}.}
\label{fig:sd-central}
\end{figure}

Jet grooming techniques have been applied to heavy-ion collisions
in order to explore whether jet quenching in the quark-gluon plasma 
modifies the hard substructure of jets
\cite{PhysRevLett.119.112301, Mehtar-Tani2017, Chang:2019nrx, Elayavalli2017, Caucal:2019uvr, Ringer_2020, Casalderrey-Solana:2019ubu, Andrews_2020, PhysRevLett.120.142302, Acharya:2019djg, Sirunyan2018}.
The large UE poses
a challenge, however, 
since fluctuations in the UE can cause groomed splittings
to be misidentified \cite{mulligan2020identifying}.
Figure \ref{fig:sd-central} shows measurements of the groomed momentum fraction, \zg{},
and the groomed jet radius, \tg{}, with the Soft Drop
grooming algorithm \cite{ALICE-PUBLIC-2020-006}. 
By using strong grooming conditions than previous measurements, the result has been
fully corrected for detector effects and background fluctuations.
We find that the \zg{} distributions in \PbPb{} collisions are consistent with those 
in \pp{} collisions, whereas a significant narrowing of the \tg{} distributions 
in \PbPb{} collisions relative to  \pp{} collisions is observed.
These measurements are compared to a variety of jet quenching models \cite{Putschke:2019yrg, LBT, Majumder_2013, Caucal:2019uvr, Caucal_2018, PhysRevLett.119.112301, Chang:2019nrx, HybridModel, HybridModelResolution, Casalderrey-Solana:2019ubu, Ringer_2020}.
All models considered are consistent with the \zg{} measurements.
Many of the models capture the narrowing effect observed in the \tg{} distributions,  
although with quantitative differences. 
This behavior is consistent with models implementing an incoherent interaction of the 
jet shower constituents with the medium, but also consistent with
medium-modified ``quark/gluon'' fractions  with fully coherent energy loss.
By isolating the theoretically well-controlled hard substructure of jets, these measurements
provide direct connection to specific jet quenching physics mechanisms, 
and offer the opportunity for future measurements to definitively disentangle them.

{
\tiny
\bibliographystyle{JHEP}
\bibliography{main.bib}
}

\end{document}